\documentclass[12pt]{article}
\usepackage{amssymb}
\newcommand{\beq}{\begin{eqnarray}}
\newcommand{\ee}{\end{eqnarray}}
\newcommand{\tr}{{\rm Tr}}

\newcommand{\De}{\Delta}

\newcommand{\email}[1]{\tt #1\vskip 7mm \rm}

\usepackage[top=2.75cm, bottom=2.75cm, left=2.75cm, right=2.75cm]{geometry}

\newcommand{\sdet}{{\rm sdet}}
\begin{document}
\begin{titlepage}
\begin{center}
\hfill ITP-UU-18 \\
\hfill YITP-SB-05-25  \\

\vskip 20mm

{\Large {\bf Ricci-flat supertwistor spaces}}
\vskip 10mm
Ulf Lindstr\"{o}m \\
Department of Theoretical Physics,\\
Uppsala University, Box 803, SE-751 08 Uppsala, Sweden\\
\email{ulf.lindstrom@teorfys.uu.se}

Martin Ro\v{c}ek \\
State University of New York\\
Stony Brook, NY 11794-3840, USA \\
\email{rocek@insti.physics.sunysb.edu}

Rikard von Unge \\
Institute for Theoretical Physics and Astrophysics\\
Faculty of Science, Masaryk University\\
Kotl\'{a}\v{r}sk\'{a} 2, CZ-611 37, Brno, Czech Republic\\
\email{unge@physics.muni.cz}

\end{center}
\vskip .2in

\begin{center} {\bf ABSTRACT } \end{center}
\begin{quotation}\noindent
We show that supertwistor spaces constructed as a K\"ahler
quotient of a hyperk\"ahler cone (HKC) with equal numbers
of bosonic and fermionic coordinates are Ricci-flat, and 
hence, Calabi-Yau. We study deformations of the 
supertwistor space induced from deformations of the 
HKC. We also discuss general infinitesimal deformations 
that preserve Ricci-flatness.
\end{quotation}
\vfill
\flushleft{\today}
\end{titlepage}
\eject
\section{Introduction and Summary}
\setcounter{equation}{0}
Supertwistor spaces \cite{f,w} have recently aroused a 
great deal of interest (\cite{w2} and citations thereof). 
However, a broader discussion of what kinds of supermanifolds
can be interpreted as supertwistor spaces does not seem to
have been given. Here we consider general supertwistor
spaces of supermanifolds whose bosonic part is
quaternion K\"ahler; such twistor spaces admit a 
canonical Einstein metric, which, as we show, is always
Ricci-flat when the number of bosonic dimensions is one less
than the number of fermionic dimensions.\footnote{For a different
approach to studying Ricci-flat supermanifolds, see \cite{rw,rw2}.} 
This is important for the recent applications of supertwistor spaces, 
which use the technology of topological strings.

Our approach depends on the Swann space \cite{s} or 
hyperk\"ahler cone (HKC) \cite{drv} of a quaternion K\"ahler
supermanifold. It allows us to study certain kinds
of new deformations of known supertwistor spaces that 
could be related to gauge theories other than $N=4$ super Yang-Mills
theory in the conformal phase. Other deformations, 
involving non-anticommutativity \cite{kz}, or orbifolding
\cite{gkrrtz,pr} do not appear to fit into our framework.

The plan of the paper is as follows: In section 2, we prove
that a supertwistor space constructed as a K\"ahler quotient
of an HKC with equal numbers of bosonic and fermionic dimensions
is super-Ricci-flat. In section 3 we show that $\mathbb{CP}^{3|4}$ \cite{w2}
and the quadric in $\mathbb{CP}^{3|3}\times\mathbb{CP}^{3|3}$ \cite{w}
arise naturally in our framework--indeed they are the supertwistor spaces
on the first two orthogonal Wolf spaces (see, {\it e.g.,} \cite{arv}). In section
4, we discuss general infinitesimal deformations of $\mathbb{CP}^{n-1|n}$,
and consider a few examples (this discussion is not based on the HKC).
In section 5, we discuss examples of deformations of the HKC. We end with
some comments and discussion.

\section{HKC's, twistor spaces, and Calabi-Yau's}
The Swann space \cite{s} or hyperk\"ahler cone (HKC) \cite{drv}
is a hyperk\"ahler variety that can be constructed from any quaternion
K\"ahler manifold (QK). The HKC has a homothety and an
$SU(2)$ isometry that rotates the hyperk\"ahler structure. The twistor space
$Z$ of the QK manifold is the K\"ahler quotient of the HKC with respect to
an arbitrary $U(1)$ subgroup of the $SU(2)$ isometry; this quotient 
is an $S^2$ bundle over the QK, where the $S^2$ parametrizes
the choices of the $U(1)$ subgroup used in the quotient.

The twistor space of a quaternion K\"ahler manifold is always
an Einstein manifold. Here we show, using the results of \cite{drv},
that if the HKC is a super-variety with an equal number of bosonic
and fermionic coordinates, then the twistor space is Ricci-flat, and hence,
Calabi-Yau.

The K\"ahler potential of an HKC can be written as:
\beq
\chi = e^{z+\bar z+K}~,
\ee
where $K$ is the K\"ahler potential of the twistor space.
The resulting metric is
\beq
g_{a{\bar b}}=\partial_a\partial_{\bar b}\chi=\chi
\pmatrix{ K_{i{\bar \jmath}}+K_iK_{\bar \jmath}
 & K_i \cr \noalign{\vskip 3mm} K_{\bar \jmath}  & 1}\ .
\ee
As the HKC is hyperk\"ahler, the determinant of this metric is 1 (up to
holomorphic diffeomorphisms). However,
\beq
\label{dets}
\det (g_{a{\bar b}})=\chi^{2n}\,\det (K_{i{\bar \jmath}})\ ,
\ee
where $K_{i{\bar \jmath}}$ is the K\"ahler metric on the twistor space,
and $2n$ is the complex dimension of the HKC. Thus, up to holomorphic
diffeomorphisms and K\"ahler transformations,
\beq
\label{nodets}
\det (K_{i{\bar \jmath}}) = e^{-2nK}~.
\ee
This means that the twistor space is Einstein with cosmological constant $2n$.
If the HKC is a {\em super}-variety with complex dimension $(2n|2m)$, then
(\ref{dets}) becomes:
\beq
\label{sdets}
\sdet (g_{a{\bar b}})=\chi^{2(n-m)}\,\sdet (K_{i{\bar \jmath}})\ ,
\ee
where $\sdet{}$ is the {\em super}-determinant. The HKC super-variety is
Ricci flat, and hence the superdeterminant in (\ref{sdets}) can be chosen to be
1 as before. Hence
\beq
\label{nosdets}
\sdet (K_{i{\bar \jmath}}) = e^{-2(n-m)K}~,
\ee
and for $n=m$, the twistor space is Ricci-flat and thus Calabi-Yau. 
Note that since it is constructed from the HKC by a K\"ahler quotient, 
it has one less complex bosonic dimension than the HKC, and thus 
its complex dimension is $(2n-1,2n)$.

Since the techniques for constructing HKC's are rather general, 
we can thus find many Calabi-Yau twistor supermanifolds. 
We now turn to some specific examples.

\section{Examples}

The simplest example of our construction arises for the supertwistor space 
$\mathbb{CP}^{3|4}$ \cite{w2}. In this case, the HKC is simply the flat 
space $\mathbb{C}^{4|4}$, and the QK manifold is the supersphere $S^{4|4}$ 
(with 4 real bosonic dimensions and 4 complex fermionic dimensions).

The next example of an HKC that we consider is a hyperk\"ahler quotient \cite{hkq} of 
$\mathbb{C}^{8|6}$ by $U(1)$; a nice feature of this example is that the Wick-rotated
version admits interesting deformations analogous to those discussed in \cite{arv2}. 
Since it is bit less obvious, we'll write this example out. We label the coordinates of
$\mathbb{C}^{8|6}\equiv\mathbb{C}^{4|3}_+\times\mathbb{C}^{4|3}_-$ as 
$z_\pm,\psi_\pm$. The HKC is the variety defined by restriction of the K\"ahler quotient 
to the quadric:
\beq
\label{hatact}
\frac{\partial}{\partial V}\hat\chi &= &0~,~~~z_+z_-+\psi_+\psi_-~=~0~,~~{\rm where}
\nonumber\\
\hat\chi &=& e^V(z_+\bar z_++\psi_+\bar\psi_+)+e^{-V}(\bar z_-z_-+\bar\psi_-\psi_-)~.
\ee
This gives
\beq
\label{HKCact}
\chi=2\sqrt{(z_+\bar z_++\psi_+\bar\psi_+)(\bar z_-z_-+\bar\psi_-\psi_-)}~,
\ee
where we may choose a gauge, {\it e.g.}, $z_+^0=-1$ and solve the constraint by
\beq
z_-^0=\sum_i^3 z_+^iz_-^i+\sum_a^4\psi_+^a\psi_-^a~.
\ee
We find the twistor space by taking a $U(1)$ K\"ahler quotient of this space; it is
actually clearer to start with $\hat\chi$; the gauged action is
\beq
\widehat{\tilde\chi}= e^{\tilde V}\left[e^V(z_+\bar z_++\psi_+\bar\psi_+)+
e^{-V}(\bar z_-z_-+\bar\psi_-\psi_-)\right]-\tilde V~;
\ee
changing variables to $V_\pm=\tilde V \pm V$, we have
\beq
\widehat{\tilde\chi}= e^{V_+}(z_+\bar z_++\psi_+\bar\psi_+)+
e^{V_-}(\bar z_-z_-+\bar\psi_-\psi_-)-\frac{(V_++V_-)}2~,
\ee
which clearly gives the K\"ahler quotient space 
$\mathbb{CP}^{3|3}\times\mathbb{CP}^{3|3}$; thus the twistor space is
the quadric given by $z_+z_-+\psi_+\psi_-=0$ (in homogeneous coordinates) 
in $\mathbb{CP}^{3|3}\times\mathbb{CP}^{3|3}$. In this construction, it is clear
that the symmetry is $SU(4|3)$. This space has been proposed 
as an alternate twistor space ({\it ambitwistor space}) relevant 
to $N=4$ super Yang-Mills theory \cite{w,w2,ww}; some recent articles that work with ambitwistors include \cite{nv,sv,ps,ms}. It is interesting to study
the QK space below this twistor space. It is the supermanifold (with 6 complex fermions!)
extension of the QK Wolf-space $\frac{SU(4)}{SU(2)\times U(2)}$. 

\section{Infinitesimal deformations of $\mathbb{CP}^{n-1|n}$}
Since the supertwistor space $\mathbb{CP}^{3|4}$ has played
such an important role in our understanding of $N=4$ super Yang-Mills theory,
it is natural to study deformations that preserve the 
Ricci flatness of the supermetric.
We now describe such general linearized deformations of
$\mathbb{CP}^{n-1|n}$ for all $n$.
Consider the ansatz for the K\"ahler potential
\beq\label{eq:pert}
K = \ln(1+z\bar z + \psi \bar \psi) + \De = K_0 + \De~,
\ee
where $\De$ should be thought of as ``small''. We calculate the different blocks 
of the metric:
\beq
A &\equiv& K_{z\bar z}\\
B &\equiv& K_{z\bar\psi}\\
C &\equiv& K_{\psi\bar z}\\
D &\equiv& K_{\psi\bar\psi}~.
\ee
The Ricci tensor is given by the complex Hessian of the logarithm of the superdeterminant
\beq
{\rm sdet}(\partial\bar\partial K)=\frac{\det(A-BD^{-1}C)}{\det D}~,
\ee
so the condition that the manifold is Ricci-flat implies (up to holomorphic diffeomorphisms) 
that the superdeterminant is a constant. We first check this for $\De =0$:
\beq
A_0 &=& e^{-K_0}\left( 1 - \bar z \otimes z e^{-K_0}\right)\\
B_0 &=&  -e^{-2K_0}\bar{z}\otimes\psi\\
C_0 &=&  -e^{-2K_0}\bar\psi \otimes z\\
D_0 &=& e^{-K_0}\left(1- \bar\psi\otimes\psi e^{-K_0}\right)~.
\ee
{}From these expressions we find
\beq
D_0^{-1} &=& e^{K_0}\left(1+\frac{\bar\psi\otimes\psi}{1+z\bar z}\right)\\
A_0-B_0D_0^{-1}C_0 &=& e^{-K_0}\left(1-\frac{\bar z\otimes z}{1+z\bar z}\right)\\
\left(A_0-B_0D_0^{-1}C_0\right)^{-1} &=&
e^{K_0}\left(1+\bar z \otimes z\right)~,
\ee
which immediately leads to the bosonic determinant
\beq
\det \left(A_0-B_0D_0^{-1}C_0\right) = e^{(1-n)K_0}(1+z\bar z)^{-1}~.
\ee
The fermionic determinant is slightly more complicated
\beq
\det{D_0} = e^{-nK_0}e^{\tr\ln(1-\bar\psi\otimes\psi e^{-K_0})}
 = e^{(1-n)K_0}(1+z\bar z)^{-1}~.
\ee
so we see that ${\rm sdet}(\partial\bar\partial K)=1$.

We now repeat the calculation for a small but nonzero $\De$.
Then the superdeterminant should be calculated with $A=A_0+A_1,B=B_0+B_1,\ldots$
where $A_1,B_1,C_1,D_1$ come entirely from $\De$. The condition that the 
superdeterminant is one implies: 
\beq
\det D = \det \left(A-BD^{-1}C\right)~,
\ee
which, to lowest order in the small $\De$ becomes
\beq
\tr D_0^{-1}D_1&\!\!=\!\!&
\tr \left[\left(A_0-B_0D_0^{-1}C_0\right)^{-1}\right.\\
&&~~~~\left.\times \left(A_1-B_1D_0^{-1}C_0-B_0D_0^{-1}C_1+B_0D_0^{-1}D_1D_0^{-1}C_0\right)
\right]~.\nonumber
\ee
This can be simplified by inserting the explicit expressions for the
unperturbed matrices:
\beq\label{eq:diff}
\tr D_1  =
\tr A_1 +zA_1\bar z + zB_1\bar \psi
+\psi C_1 \bar z + \psi D_1 \bar\psi~.
\ee

An example of a solution to this equation is found in the next section.
Since the full bosonic $SU(n)$ symmetry is nonlinearly realized, 
all nontrivial deformations appear to necessarily break it, and it is 
unclear what if any applications for Yang-Mills theories such
deformations may have.\footnote{The most general ansatz for an infinitesimal 
perturbation that {\em does} preserve $SU(n)$ symmetry has the form 
$$\Delta=\sum_{i=1}^n\sum_{j_1\dots j_i\bar k_1\dots \bar 
k_i=1}^na^i_{j_1\dots j_i\bar k_1\dots \bar k_i}\frac{\psi^{j_1}\dots\psi^{j_i} 
\bar\psi^{k_1}\dots\bar\psi^{k_i}}{(1+z\cdot\bar z+\psi\cdot\bar\psi)^i}~,$$ 
where the $a^i$ are hermitian arrays $a^i_{j_1\dots j_i\bar k_1\dots \bar k_i}
=\bar a^i_{\bar k_1\dots \bar k_i j_1\dots j_i}$; we have checked that for $n=3$,
eq.~(\ref{eq:diff}) implies $\Delta=\sum_{ij}\,(a_{ij}\psi^i\bar\psi^j)
/(1+z\cdot\bar z+\psi\cdot\bar\psi)$, which can be absorbed by a 
holomorphic change of coordinates $\psi\to\psi+\frac12 a_{ji}\psi^j$.} An intriguing 
possibility is that some such deformation could describe the nonconformal phase
of $N=4$ super Yang-Mills theory.\footnote{We thank R.~Wimmer for this suggestion.}

\section{Examples of deformations of the HKC}

HKC's that are constructed as hyperk\"ahler quotients can be deformed by
deforming the quotient (see, {\it e.g.}, \cite{arv2}). Both the examples described
in section 3 can be deformed; there are many deformations of the quadric in 
$\mathbb{CP}^{3|3}\times\mathbb{CP}^{3|3}$ (ambitwistor space), and we begin
a few examples. 

\subsection{Some deformations of ambitwistor space}
The simplest kind of deformations involve modifying the 
$U(1)$ charges of the fermions when we perform the hyperk\"ahler quotient,
that is (\ref{hatact}). Instead of the diagonal $U(1)$, we may use any subgroup
of the $U(3)$ that acts on the fermion hypermultiplets. For example, we may
choose a $U(1)$ that acts only on one or two pairs of the fermions; if we label
the fermionic coordinates as $\psi_\pm,\psi_0$ according to their $U(1)$ charge,
(\ref{hatact}) becomes
\beq
\label{hatactdef}
\frac{\partial}{\partial V}\hat\chi &= &0~,~~~z_+z_-+\psi_+\psi_-~=~0~,~~{\rm where}
\nonumber\\
\hat\chi &=& e^V(z_+\bar z_++\psi_+\bar\psi_+)
+e^{-V}(\bar z_-z_-+\bar\psi_-\psi_-) +\psi_0\bar\psi_0~.
\ee
This gives
\beq
\label{HKCactdef}
\chi=2\sqrt{(z_+\bar z_++\psi_+\bar\psi_+)(\bar z_-z_-+\bar\psi_-\psi_-)} +\psi_0\bar\psi_0~.
\ee
We find the twistor space by taking a $U(1)$ K\"ahler quotient of this space; it is
actually clearer to start with $\hat\chi$; the gauged action is
\beq
\widehat{\tilde\chi}= e^{\tilde V}\left[e^V(z_+\bar z_++\psi_+\bar\psi_+)+
e^{-V}(\bar z_-z_-+\bar\psi_-\psi_-) +\psi_0\bar\psi_0\right]-\tilde V~.
\ee
In homogeneous coordinates, this gives the quadric 
$z_+z_-+\psi_+\psi_-=0$ in K\"ahler supermanifold with K\"ahler potential
\beq
K=\frac12\ln(M_+)+\frac12\ln(M_-)
+\ln\left({1+\frac{\psi_0\bar\psi_0}
{2\sqrt{M_+M_-}}}\right)~.
\ee
where
\beq
M_+=(z_+\bar z_++\psi_+\bar\psi_+)~~,~~~
M_-=(\bar z_-z_-+\bar\psi_-\psi_-)~.
\ee
This deformation of the ambitwistor space has the symmetry $SU(4|n)\times SU(6-2n)$,
where $n$ is the number of charged doublets $\psi_\pm$.

Other choices of the $U(1)$ action on the fermions give other deformations with different
residual symmetries. 

If we change the metric of the initial flat space whose $U(1)$ quotient gives the HKC,
{\it e.g.}, so that the global symmetry group acting on the fermions becomes $U(2,1)$,
then new deformations analogous to those discussed in \cite{arv2} become available.

Another kind of deformation arises from the observation that the underlying bosonic
quaternion K\"ahler manifold, the unitary Wolf space 
$SU(4)/[SU(2)\times SU(2)\times U(1)]$ is the same as the orthogonal Wolf space
$SO(6)/[SO(3)\times SO(3)\times SO(2)]$, and hence, as explicitly shown in \cite{arv},
the corresponding HKC's are equivalent. These deformations are rather complicated,
and lead to a variety of residual symmetry groups.

\subsection{Deformations of  $\mathbb{CP}^{3|4}$}

The chiral twistor space $\mathbb{CP}^{3|4}$ appears to 
be more useful than ambitwistor space, {\it e.g.}, \cite{w2,rsv,b,rv}.
However, in contrast to the deformations of ambitwistor space,
it is not obvious that any of the deformations of $\mathbb{CP}^{3|4}$
preserve the bosonic conformal symmetry, limiting their possible applications. 

The deformations that we consider are similar to the 
last kind mentioned for the ambitwistor space: namely,
deformations based on the identification of $S^4$ 
as both the first symplectic Wolf space $Sp(2)/[Sp(1)\times Sp(1)]$ as
well as the first orthogonal Wolf space $SO(5)/[SO(3)\times SO(3)]$. This identification
implies that the corresponding HKC's are equivalent, and this was indeed explicitly shown
in \cite{arv}. Thus $\mathbb{C}^4\equiv \mathbb{C}^{10}\!/\!/\!/\!SU(2)$, the hyperk\"ahler quotient of $\mathbb{C}^{10}$ by $SU(2)$, and, as also shown in \cite{arv}, 
it is natural to write this as $\mathbb{C}^{20}\!/\!/\!/[SU(2)\times U(1)^5]$. 
This suggests a variety of deformations that we now describe. 

Two deformations that we consider arise by coupling the $U(1)$'s and the $SU(2)$ to the
fermions as follows (we use the language of $N=1$ superspace $\sigma$-model Lagrangians)
\beq
L_{f} = 
\int\! d^4\theta\, \psi e^{\sigma V} \bar{\psi} e^{q\sum_i V_i} +
\tilde{\bar{\psi}} e^{-\sigma V} \tilde{\psi} e^{-q\sum_i V_i}
+\left[\int\! d^2\theta\,\sigma \psi \phi \tilde{\psi} + q \psi \tilde{\psi} \sum_i \phi_i + c.c.\right]~,
\nonumber\\
\ee
where $\sigma = 0,1$ and $q$ is arbitrary.

\subsection{The $U(1)$ deformation}
To begin with we set $\sigma = 0$ and study the $U(1)$ deformation.
If we integrate out the $M$'th $U(1)$ vector multiplet we get the following two
equations
\beq\label{eq:u1motion}
0 &=& z^M\tilde{z}^M + q \psi \tilde\psi \nonumber\\
0 &=& z^M e^{V}\bar{z}^M e^{V_M}
-\bar{\tilde{z}}^M e^{-V} \tilde{z}^M e^{-V_M}+
q \psi  \bar{\psi} e^{q\sum_M V_M} - q \bar{\tilde{\psi}}
 \tilde{\psi} e^{-q\sum_M V_M}~.
\ee
We introducing the notation
\beq
M_M = z^M e^{V} \bar{z}^M \;\;\;~,~~~~~
\tilde{M}_M = \tilde{\bar{z}}^M e^{-V} \tilde{z}^M~.
\ee
Though the deformation makes sense, to simplify our calculations, 
we work only to the lowest nontrivial order in $q$ throughout this subsection.
This means that we can compare directly to the results of section 4.
Then the second equation of motion simplifies to
\beq
M_M e^{V_M}-\tilde{M}_M e^{-V_M}+q\left(\psi\bar{\psi}-
\bar{\tilde{\psi}}\tilde{\psi}\right)+{\cal O}(q^2) = 0~.
\ee
Solving for $V_M$ we get
\beq
V_M = \ln\sqrt{\frac{\tilde{M}_M}{M_M}}-\frac{q(\psi\bar{\psi}-
\bar{\tilde{\psi}}\tilde{\psi})}{2\sqrt{M_M\tilde{M}_M}}
+{\cal O}(q^2)~.
\ee
Plugging this back into the action we get
\beq
\int d^4\theta \left( \sum_{M=1}^{5} 2\sqrt{M_M\tilde{M}_M}
+ \psi \bar{\psi}+\bar{\tilde{\psi}}\tilde{\psi}
+\frac{q}{2}\left(\psi\bar{\psi}-
\bar{\tilde{\psi}}\tilde{\psi}\right)
\ln\frac{\Pi_{j=N}^{5}\tilde{M}_N}
{\Pi_{K=1}^{5}M_K}
\right. \nonumber\\ \left.
+\int d^2\theta \sum_M z^M\phi \tilde{z}^M
+{\cal O}(q^2)~.
\right)
\ee
It is not possible to solve the first equation in (\ref{eq:u1motion}) in an $SU(2)$
covariant way. Rather we set
\beq
\tilde{z}^M = i\sigma_2 z^M - q \psi\tilde{\psi}\frac{\sigma_1z^M}
{(z^M\sigma_1z^M)} +{\cal O}(q^2)~.
\ee
Inserting this and keeping only terms to lowest nontrivial order in $q$ we get
\beq
\int d^4 \theta \left(\sum_M \left(
2z^Me^V \bar{z}^M + q\psi\tilde{\psi}\frac{z^M\sigma_3 e^V\bar{z}^M}{z^M\sigma_1 z^M}
+q\bar{\tilde{\psi}}\bar{\psi} \frac{z^M e^V \sigma_3 \bar{z}^M}
{\bar{z}^M\sigma_1\bar{z}^M}\right)
 +\psi\bar{\psi}+\bar{\tilde{\psi}}\tilde{\psi}
\right)\nonumber\\
+\int d^2 \theta \sum_M \left(
z^M\phi i \sigma_2 z^M - 
q \frac{z^M\phi\sigma_1 z^M}{z^M\sigma_1z^M}\psi\tilde{\psi}
\right) +{\cal O}(q^2) +c.c.
\ee
It is now straightforward to proceed to integrate out the $SU(2)$ gauge field
leading to the holomorphic constraints
\beq
\left(
\begin{array}{cc}
\sum_{M}z_+^Mz_+^M & \sum_M z_+^M z_-^M\\
\sum_M z_-^M z_+^M & \sum_M z_-^M z_-^M
\end{array}\right) =
\frac{q}{2}\psi\tilde{\psi}\left(
\begin{array}{cc}
-\sum_M\frac{z_+^M}{z_-^M} & 0\\
0 & \sum_M\frac{z_-^M}{z_+^M}
\end{array}\right)~,
\ee
together with the action
\beq
4\int &\!\!\!\!d^4\theta \sqrt{\det\sum_M\left(
\bar{z}^M\otimes z^M + \frac{q}{2} \frac{\psi\tilde{\psi}}{z^M\sigma_1z^M}
\bar{z}^M \otimes z^M \sigma_3 + \frac{q}{2}
\frac{\bar{\tilde{\psi}}\bar{\psi}}{\bar{z}^M\sigma_1\bar{z}^M}
\sigma_3 \bar{z}^M\otimes z^M
\right)}&\nonumber\\ 
&+~\psi\bar{\psi}+\bar{\tilde{\psi}}\tilde{\psi}~.\qquad\qquad\qquad\qquad\qquad
\qquad\qquad\qquad\qquad\qquad\qquad\qquad&
\ee
The holomorphic constraint can be solved as
\beq
z_\pm^M = w_\pm^M\mp e_\pm^M\frac{q}{4}\psi\tilde{\psi}\sum_N
\frac{w_\pm^N}{w_\mp^N} +{\cal O}(q^2)~,
\ee
where we have used the solutions $w_\pm$ of the $q=0$ constraints in terms
of four complex coordinates $u_i$
introduced in \cite{arv} 
\beq\label{eq:w}
w_+^M &=& \left(-2u_3 , -i\frac{u_3^2}{u_1}- iu_1 , -\frac{u_3^2}{u_1}+u_1 ,
\frac{u_3 u_4}{u_1}+u_2 , -i\frac{u_3u_4}{u_1}-iu_2\right)\nonumber\\
w_-^M &=& \left(2u_4 , i\frac{u_3u_4}{u_1}-iu_2 , \frac{u_3u_4}{u_1} - u_2 ,
-\frac{u_4^2}{u_1}+u_1 , i\frac{u_4^2}{u_1} +iu_1\right)
\ee
and the two vectors $e_+$ and $e_-$ are defined such that
\beq
e_+\cdot w_+ = e_-\cdot w_- = 1\nonumber\\
e_+\cdot e_+ = e_-\cdot e_- = e_+\cdot w_- = e_- \cdot w_+ = e_+ \cdot e_- = 0
\ee
with the explicit solutions
\beq\label{eq:e}
e_+ &=& \left(0,\frac{i}{2u_1},\frac{1}{2u_1},0,0\right)\nonumber\\
e_- &=& \left(0,0,0,\frac{1}{2u_1},-\frac{i}{2u_1}\right)
\ee

To show that this indeed gives a HKC one has to rewrite this expression in the form
\beq
e^{u+\bar{u}+K(z,\bar{z},\psi,\bar{\psi})}
\ee
The variable which we factor out and which will play the role of $u$ is $u_1$. The variables
on which $K$ depend are then the original variables rescaled by $u_1$. One can show that
$K$ will be of the form (\ref{eq:pert}) with a $\Delta$ looking like
\beq
\label{qdelta}
\Delta = \frac{f(u,\bar{u})\psi\tilde{\psi} +\bar{f}(u,\bar{u})\bar{\tilde{\psi}}\bar{\psi}}
{1+|u|^2+|\psi|^2+|\tilde{\psi}|^2}
\ee
where is $f(u,\bar{u})$ is a particular function of the bosonic variables;
(\ref{eq:diff}) implies that $f$ has to satisfy the differential equation
\beq\label{eq:fdiff}
\frac{\partial^2}{\partial u^i \partial \bar{u}^i} f+
u^i \bar{u}^k \frac{\partial^2}{\partial u^i \partial \bar{u}^k} f
-u^i\frac{\partial}{\partial u^i}f
+\bar{u}^i\frac{\partial}{\partial \bar{u}^i}f
-f =0~.
\ee
A straightforward but lengthy calculation confirms that the particular $f$ 
one gets from the $U(1)$ deformation does indeed satisfy (\ref{eq:fdiff}).\footnote{
Though (\ref{qdelta}) resembles the $SU(4)$ invariant ansatz noted in section 4,
there is no choice of $f(u,\bar u)$ which is consistent with the symmetry. }

\subsection{$SU(2)$ deformations}
In this section we put $q=0$. Then the holomophic constraints from
integrating out the $M$'th $U(1)$ is
\beq
 z_{Ma}\tilde{z}_M^a = 0
\ee
This is solved by putting $\tilde{z}_M^a =
\left(i\sigma_2\right)^{ab}z_{bM}$. To get a more unified notation, let us
also introduce $\psi_a = \frac{1}{\sqrt{2}}(\psi_{1a}-i\psi_{2a})$, $\tilde{\psi}^a =
\frac{1}{\sqrt{2}}
\left(i\sigma_2\right)^{ab} (\psi_{1b}+i\psi_{2b})$. Thus, integrating out the $U(1)$'s
and using this new notation, the bosonic piece of the action becomes
\beq
L_b = \sum_M\left(\int d^4\theta
z_M e^V \bar{z}_M +\left[\int d^2 \theta
z_M \phi i\sigma_2 z_M + c.c.\right]
\right)
\ee
and the fermionic piece becomes
\beq
L_f = i\int d^4\theta \psi_1 e^V \bar{\psi}_2
-\psi_2 e^V \bar{\psi}_1+
\left[\frac i2 \int d^2\theta
\psi_1 \phi i\sigma_2 \psi_2 - \psi_2 \phi i\sigma_2 \psi_1
+c.c. \right]
\ee

Integrating out the $SU(2)$ gauge field, and letting the $SU(2)$ indices
run over $+$ and $-$, the holomorphic constraints can be written as
\beq
 z_+\cdot z_+ + 2\psi_+^1\psi_+^2 &=&\nonumber\\
 z_+\cdot z_- +\psi_+^1\psi_-^2+\psi_-^1\psi_+^2 &=& 0\\
 z_-\cdot z_- + 2\psi_-^1\psi_-^2 &=&\nonumber
\ee
Again using the variables (\ref{eq:w}) and (\ref{eq:e})
we were able to write down an explicit solution to the holomorphic constraints
\beq
z_+^M &=& w_+^M 
-e_+^M\psi_+^1\psi_+^2 - e_-^M\psi_-^1\psi_+^2\nonumber\\
z_-^M &=& w_-^M
-e_-^M\psi_-^1\psi_-^2 -e_+^M\psi_+^1\psi_-^2
\ee

Now we turn to integrating out $V$. This can be done with the result
\beq
\sqrt{\det\left(\bar{z}\cdot z -
\bar{\psi}_1\psi_2+\bar{\psi}_2\psi_1\right)}
\ee
where the determinant is taken over the $+$ and $-$ indices. In this
expression one furthermore has to use the expressions of $z$ in terms of
the $u$ variables given above. 
By introducing supercoordinates
\beq
\phi^A_\pm = \left(
\begin{array}{c}
z^M_\pm \\ \psi^i_\pm
\end{array}\right)
\ee
and the supermetric
\beq
g_{AB} = \left(\begin{array}{cc}
\delta_{MN} & 0\\
0 & \epsilon_{ik}
\end{array}\right)
\ee
one may rewrite the $SU(2)$ matrix over which we are taking the determinant in a manifestly $OSp(5|2)$ covariant way
\beq
\left(\begin{array}{cc}
\phi^A_+ g_{AB} \phi^B_+ & \phi^A_+ g_{AB} \phi^B_-\\
\phi^A_- g_{AB} \phi^B_+ & \phi^A_- g_{AB} \phi^B_-
\end{array}\right)
\ee
In the case without fermions it was shown in \cite{arv} that the manifest 
$O(5)$ symmetry is enhanced to an $SU(4)$ symmetry. We do not know 
if this occurs in the presence of the fermions; one might hope that the 
symmetry is enhanced to the supergroup $SU(4|2)$, which
is the superconformal group of of ${\cal N} =2$ supersymmetric $SU(N_c)$
Yang-Mills theory with $2N_c$ hypermultiplets in the fundamental 
representation. We have not found evidence one way or the other.

To show that this indeed gives a HKC we again have to rewrite this expression in the form
\beq
e^{u+\bar{u}+K(z,\bar{z},\psi,\bar{\psi})}
\ee
Again the variable that we choose to factor out is $u_1$ and the 
remaining variables of the K\"ahler potential are the original 
variables rescaled by $u_1$. We thus find a compact form for 
the K\"ahler potential of the twistor space. It is however not
of the form (\ref{eq:pert}) since the perturbation in no way 
can be thought of as being ``small".

\section{Conclusions}
We have shown that supertwistor spaces constructed from hyperk\"ahler cones
with equal numbers of bosonic and fermionic coordinates are super-Ricci-flat.
We have used this result to discuss deformations of supertwistor spaces. All the deformations
of the chiral supertwistor space $\mathbb{CP}^{3|4}$ that we found appear to break 
conformal invariance (though one case we discussed is unclear), whereas it is simple
to deform ambitwistor space (the quadric in $\mathbb{CP}^{3|3}\times\mathbb{CP}^{3|3}$)
in ways that preserve a variety of superconformal symmetries. It would be interesting to
study if any of these deformations arise in superconformal Yang-Mills theories.

\section*{Acknowledgments:}
UL and RvU are grateful for the stimulating atmosphere
at the Second and Third Simons Workshop in 
Mathematics and Physics at the C.N. Yang
Institute for Theoretical Physics, Stony Brook, where this work was
completed. MR is happy to thank R. Wimmer for comments.
UL acknowledges support in part by EU contract
HPNR-CT-2000-0122 and by VR grant 650-1998368. 
The work of MR was supported in part by 
NSF Grant No. PHY-0354776 and Supplement for
International Cooperation with Central and 
Eastern Euorpe PHY 0300634.  The work of RvU was 
supported by the Czech Ministry of Education under
the project MSM 0021622409 and ME649.

\end{document}